\documentclass[traditabstract]{aa}
\usepackage[authoryear]{natbib}
\usepackage{graphicx}
\usepackage{amsfonts}
\def\lunits{$\rm erg\,s^{-1}$~}

\def\cunits{$\rm cm^{-2}~$}
\def\sax{{\it BeppoSAX~}}
\def\xmm{{\it XMM-Newton~}}
\def\chandra{{\it Chandra~}}

\begin{document}
\title{Molecular lines as tracers of Compton-thick AGN ?}


  \titlerunning{Molecular lines as tracers of Compton-thick AGN}
    \authorrunning{I. Georgantopoulos et al.}

   \author{I. Georgantopoulos\inst{1,2}
           E. Rovilos \inst{3}
           A. Akylas \inst{2}
           E. Xilouris \inst{2}
                     }

   \offprints{I. Georgantopoulos, \email{ioannis.georgantopoulos@oabo.inaf.it}}

   \institute{INAF-Osservatorio Astronomico di Bologna, Via Ranzani 1, 40127,
              Bologna, Italy 
              \and 
              Institute of Astronomy and Astrophysics, National Observatory of
              Athens, I. Metaxa \& V. Pavlou str., Palaia Penteli, 15236 Athens,
              Greece
              \and
              Max-Planck-Institut f\"{u}r Extraterrestrische Physik,
              Giessenbachstra\ss e, 85748 Garching, Germany
             }

   \date{Received ; accepted }

\abstract{Recently, Papadopoulos et al., 2010 using sub-mm CO molecular line
          observations of nearby ultra-luminous IRAS galaxies,  (U)LIRGs with
          $\rm L_{IR}\geq 10^{11}\,L_{\odot}$,  have found that exceptionally
          large gas column densities ($\rm N_H\geq 10^{25}\,cm^{-2}$) can
          be present across some of the very dense gaseous disks that are typically
          found in these objects. They also proposed a
          diagnostic for finding such sources using CO and HCN molecular lines.
          Given that such high column densities are expected to absorb any
          X-ray luminous AGN, yielding Compton-thick sources, we set out to
          explore whether this can be discerned using X-ray observations. More
          specifically we examine X-ray spectral observations of 14 sources in
          their sample, using public \chandra observations (0.5-10\,keV) for eleven sources 
           as well as \sax results (2-100\,keV) from the literature for another three sources. 
            Our  goal is to find candidate Compton-thick AGN
          and to check whether the molecular line selection criterion is 
           successful in selecting such systems. 
          X-ray spectroscopy reveals four candidate Compton-thick AGN of which half fall within the
          high obscuration region in the molecular line ratio diagnostics. Of
          the remaining five sources falling into the `high dust obscuration'
          box, one (Mrk273) is highly obscured
          ($\rm N_H\sim 4\times 10^{23}\,cm^{-2}$) while in the other four the
          X-ray emission is most probably associated with star-forming
          processes rather than an AGN on the basis of their X-ray and
          mid-infrared properties. Overall, we argue that although this method
          as expected cannot  recover all Compton-thick AGN, there are no examples of
          X-ray luminous AGN inside that region that have low obscuration, suggesting
          that this method
          is efficient in finding heavily obscured AGN in dust-enshrouded
          star-forming galaxies. The above results bear important implications for 
          future joint ALMA and
          X-ray observations for the detection of Compton-thick AGN.
     \keywords {X-rays: general; X-rays: diffuse background; X-rays: galaxies;
         Galaxies: active; Submillimeter: galaxies; Infrared: galaxies}}
   \maketitle
%

\section{Introduction} 
X-ray surveys provide the most efficient tool for identifying AGN
\citep{Brandt2005}. This is because the hard X-rays can penetrate large amounts
of dust and gas and thus can reveal hidden AGN which would be very difficult to
detect at optical wavelengths. However, even the hardest X-rays cannot easily
escape from very large absorbing column densities, $>10^{24}$\,\cunits,
\citep{Murphy2009}. A column density of a few times $10^{24}$\,\cunits absorbs
X-rays up to 20\,keV, while for column densities higher than $10^{25}$\,\cunits
practically all X-ray photons are attenuated. Such immense column densities
($>10^{24}$\,\cunits) correspond to the so-called Compton-thick AGN, where
Compton scattering becomes the dominant attenuation process. In these cases,
reflected emission from the backside of the obscuring screen, in a form of
a torus, remains the only evidence of the AGN presence in X-rays
\citep{Matt2004}. 
 
Such extreme column densities could be relatively easily explored in the IR or
sub-mm part of the spectrum, through their thermal or molecular emission.
For example, sub-mm observations have drawn much attention recently as a means
of discovering heavily obscured AGN \citep[e.g.][]{Alexander2005}. The discovery
of significant dust optical depths even at submillimeter wavelengths
\citep[in the case of Arp220;][]{Sakamoto2008} reveals that such highly
obscured systems may not be uncommon. Recently, \citet{Papadopoulos2010}
(hereafter P10) made the interesting suggestion that observations of suitable
molecular lines can reveal cases of very high column densities (and thus
possible obscuration) present in some reservoirs of molecular gas in LIRGs. The
average  molecular gas densities in some of these systems can be so high
that if the  gas is distributed in a gaseous disk it can yield extremely high
column densities and dust optical depths that suppress the high-J (6-5) CO line
emission to levels well below those expected in starburst galaxies such as
ULIRGs. In such cases, if an active galactic nucleus were embedded in the
(U)LIRG, the line of sight towards it would also be severely obscured making
this source Compton-thick. A molecular line diagnostic to identify such
severely obscured sources was proposed by P10. This helps distinguish
between the degenerate possibilities of high dust obscuration and large
reservoirs of non star-forming cool and diffuse gas, whose high-J (6-5) CO
lines are genuinely faint rather than immersed to a strong dust continuum.
They suggest that high fractions of very dense molecular gas (with
$\rm n(H2)>10^4\,cm^{-3}$) can be traced by a high HCN(1-0)/CO(1-0) ratio (R$>$0.1). 
  Then a low ($\rm R<0.3$) CO(6-5)/(3-2) ratio  
   can only be the result of high dust optical depths ($\tau_{434 \mu m}\sim 1$). 
    Consequently there should be a very
high probability of Compton-thick AGN for (U)LIRG hosts lying inside their
aforementioned selection ``window''.  On the other hand, we note that gaseous
disks (in the center of which the AGN are presumably embedded) with low average
column densities can often yield large column densities along the particular
line of sight towards the AGN as a result of their highly turbulent nature
\citep*{Wada2009}. Thus Compton-thick sources can be expected also outside the
dust-obscuration selection window, the vast majority of sources inside it
however should be Compton thick. In other words, if molecular gas disks have on
average high column densities, as the supression of CO J=6-5 emission from
their star-forming gas would imply, this will almost certainly be the case for
the special line of sight towards the AGN as well, but the reverse does not
necessarily hold. Thus the selection window of highly dust obscured
star-forming systems advocated by P10 ought to select almost exclusively
Compton-thick AGN, but not all of them.

In this paper we are examining the X-ray properties of the systems in P10,
using both literature spectra as well as \chandra archival observations.
Our goal is to test whether AGN residing in the P10 box have a higher
probability for being classified as Compton-thick sources in X-rays. 
We adopt $\rm H_o= 75\,km\,s^{-1}\,Mpc^{-1}$, $\rm \Omega_{M}=0.3$,
$\rm \Omega_\Lambda=0.7$ throughout the paper.

\section{Data \& analysis }
\subsection{The sample} 
P10 present molecular line spectra for 20 (U)LIRGs. For six objects there are
no HCN observations available so that the HCN/CO vs. CO(6-5)/CO(3-2) diagnostic
cannot be applied. For the remaining 14 objects there are X-ray spectra
available either in the literature or in publicly available \chandra
observations. Although in many cases the \chandra observations may have
been presented elsewhere, we chose to re-analyze the data in order to provide a
uniform analysis of the sample. To our knowledge, in six cases X-ray spectral fits to the 
\chandra data are presented here for the first time: VII\,Zw\,031, IRAS\,10565+2448, 
IRAS\,12112+0305, Arp\,193, IRAS\,17208-0014, IRAS\,23365+3604. 
 However, X-ray luminosities or hardness ratios derived from \chandra
  observations  have been presented before in the literature 
   for some of those (see Table\,\ref{xrayprop}). 
In the cases of Mrk\,231, NGC\,6240 and NGC\,7469, where there are 
 high signal-to-noise spectra 
 in the $2-100$\,keV band reported in the literature from {\it BeppoSAX}, 
 we prefer to present that data instead, as high energy observations can be much
more efficient in detecting Compton-thick sources. 

\subsection{Spectral fits} 
We fit the {\it Chandra} data in the 0.5-10 keV band using {\sl XSPEC} v.12.6 \citep{Arnaud1996}. 
 Initially, we fit
an absorbed  power-law model to the spectra. For more complex spectra,
 and judging on the basis of the data residuals from the model fit,
 we additionally fit a thermal component (Raymond-Smith) to account for X-ray
emission from  star-forming processes, and/or an FeK$\alpha$ line around
6.4\,keV.
  The C-statistic has been employed \citep{Cash1979}. The
errors correspond to the 90\% confidence level as is customary in X-ray
Astronomy. The X-ray spectral properties and luminosities are presented in
Table\,\ref{xrayprop}. In the \chandra data presented here, an FeK$\alpha$ line has been detected 
in the case of the sources IRAS\,09320+6134 (see discussion below) and the Seyfert-2 galaxy 
Mrk\,273.  An  in-depth presentation of the 
 \chandra spectrum of this source can be found in \citet{Xia2002} and
\citet{Ptak2003}. 

We use the following criteria to classify a source as Compton-thick: a) the detection  of an 
 absorption turnover b) a flat X-ray spectrum 
 with $\Gamma \sim 1$ or flatter, suggestive of a reflection-dominted spectrum, 
  or finally c) a high EW ($\rm >0.8\,keV$) FeK$\alpha$ line, indicating a strongly 
   absorbed continuum
\citep[see discussion in ][for justification and explanation of these diagnostics]{Matt2004,Akylas2009,Georgantopoulos2009}
On the basis of the above criteria, we consider four Compton-thick sources:
IRAS\,09320+6134, Mrk\,231, Arp\,220, and NGC\,6240. In
the cases of Mrk\,231 and NGC\,6240 the absorption turnovers have been directly
detected by {\it BeppoSAX} (\citealt{Braito2004} and \citealt{Vignati1999}
respectively) and therefore these can be considered as bona-fide Compton-thick sources.
 In the case of Arp\,220, the spectrum
is flat ($\Gamma \sim 1.1$) but no Fe line can be detected in the \chandra data. However, in
the \xmm  data \citep{Iwasawa2005} a high EW (1.9\,keV) FeK$\alpha$ line at 6.7\,keV 
 has been clearly detected rendering this source a highly probable Compton-thick source. 
  The case of IRAS\,09320+6134 (UGC\,5101) is the most ambiguous one; 
   we detect a flat spectrum ($\Gamma \sim 1.2$), albeit  with a large uncertainty, together with a high
  EW ($\rm \sim 3.6\,keV$) FeK$\alpha$ line. 
  The combination of a flat spectrum with a high EW FeK$\alpha$ line 
      rather points towards a Compton-thick  AGN. 
  The EW derived here is consistent with the EW obtained by \citet{Ptak2003}. 
   \citet{Imanishi2003} analyse both the \xmm and the \chandra data finding 
    a considerably lower EW ($\rm \sim 400\,eV$). Finally,
\citet{Gonzalez-Martin2009} 
     claim the detection of a mildly Compton-thick AGN with a column density of $\rm N_H\approx 1.4^{+0.4}_{-0.04}\times 10^{24}$\,\cunits.  Finally, in the case of IRAS\,23365+3604, the spectrum is flat but with a large uncertainty. Moreover, there is no 
      evidence for a high-EW FeK$\alpha$ line, therefore the possibility that this 
      source is Compton-thick is weaker. 
    
 \begin{table*}
\centering
\caption{X-ray Properties}
\label{xrayprop}
\begin{tabular}{ccccccccccc}
\hline\hline
Name                    & ObsID & $\rm N_H$              & $\Gamma$               & kT   & E  & EW  & $ \rm logL_X$ & $\rm logL_{FIR}$ & class. & Ref. \\
 (1)                    & (2)    & (3)                    & (4)                    & (5)      & (6)       & (7)  & (8) & (9) & (10) & (11) \\
\hline
VII Zw031               &   7887     & $<0.12$                & $1.92^{+0.47}_{-0.32}$ & -   & &        &         41.38           & 11.71              &    - &  x \\
05189-2524          &    2034    & $7.6^{+0.85}_{-0.94}$  & $1.75^{+0.18}_{-0.23}$ & $0.07_{-0.11}^{+0.07}$ & 
 & &   43.30           & 11.83              &   Sy2 &  x,a  \\
Arp55                   &    6857    & $<0.92$                & $2.65_{-0.83}^{+5.3}$  & -  & &    &  40.10           & 11.48              &    - & x,b  \\ 
09320+6134$\dagger$ &     2033   & $<0.3$                     & $1.17_{-0.21}^{+0.57}$ & - & $6.76\pm0.04$ & $3500^{+1500}_{-1300}$ &   41.56           & 11.74              &  L&    x,a,b  \\ 
10565+2448          &   3952     & $0.18_{-0.08}^{+0.08}$ & $2.13^{-0.24}_{+0.24}$ & -      & &   &              41.19           & 11.78              &   - & x,d   \\
12112+0305          &    4934    &  $<1.2$                     & $2.78^{+1.21}_{-0.99}$ & -                & &      & 40.36           & 12.09              &    L & x,d  \\ 
Mrk231$\dagger$         & -      & $273^{+100}_{-68}$   & $2.95^{+0.15}_{-0.16}$ &  & $6.39^{+0.17}_{-0.15}$ &
 $180^{+170}_{-150}$ &    42.41           & 12.15              &     QSO & e,h,a \\
Arp193                  &   7811     & $0.10^{+0.08}_{-0.10}$ & $1.71^{+0.38}_{-0.36}$ & -         & &   &             40.99           & 11.41              &  - &   x  \\
Mrk273                  &  809      & $46.7_{-5.4}^{+7.0}$   & 2.                     & $0.96^{+0.05}_{-0.07}$ &  $6.59\pm 0.04$ & $220^{+120}_{-100}$  & 42.99           & 11.93              &    Sy2 & x,f,a  \\ 
Arp220$\dagger$         & 869       & $0.43^{+0.54}_{-0.28}$ & $1.11_{-0.28}^{+0.52}$ & $0.92_{-0.05}^{+0.15}$ & & &  40.95           & 11.96              &  L &   x,a,j,k   \\ 
NGC6240$\dagger$        & -      & $179^{+40}_{-15}$                  & $1.98^{+0.25}_{-0.16}$  &        &    $6.50^{+0.15}_{-0.07}$ & $1880^{+1590}_{-730}$ &  42.01           & 11.57    &  L &        h,g,a \\ 
17208-0014          &    2035    & $0.13_{-0.09}^{+0.10}$ & $1.66_{-0.30}^{+0.26}$ & -            &   &       &  41.28           & 12.17              & HII & x     \\
NGC7469                 & -      &    $<0.03$   &  $2.05^{+0.02}_{-0.03}$ &     &  $6.35^{+0.14}_{-0.07}$ & $84^{+20}_{-19}$    &  43.27       & 11.30 &  Sy1.5 & h    \\
23365+3604          &    4115    & $<0.4$                     & $1.2^{+0.47}_{-0.43}$  & -          &  &       &      41.38           & 11.96              &    L & x,i  \\
 \hline\hline
\end{tabular}
\begin{list}{}{}
\item The columns are:
      (1) Name; A  $\dagger$ denotes candidate Compton-thick sources 
      (2) \chandra Observation ID in the case where we performed our own spectral fits
      (3) Column density in units of $10^{22}$\,\cunits
      (4) Power-law photon index 
      (5) Temperature of the thermal component in keV
      (6) Energy of the FeK$\alpha$ line in keV 
      (7) EW of the FeK$\alpha$ line in eV 
      (8) Logarithm of the {\it absorbed} hard (2-10\,keV) X-ray luminosity (\lunits)
      (9) Logarithm of the far-infrared luminosity (40-100\,$\rm \mu m$) in units
          of solar luminosity
      (10) Optical classification according to \citet*{Veilleux1999} or the NASA Extragalactic Database 
      (11) X-ray reference: a. \citet{Ptak2003}, b. \citet{Gonzalez-Martin2009},
      c. \citet{Imanishi2003} d. \citet{Iwasawa2009} e. \citet{Braito2004}
      f. \citet{Xia2002} g. \citet{Vignati1999} h. \citet{Dadina2007} i. \citet{Teng2005} j. \citet{Clements2002} k. \citet{Iwasawa2005} 
       x. this work.   
   \end{list}
\end{table*}

\subsection{Galaxies or AGN ?}
\begin{figure}
\begin{center}
\includegraphics[width=9.cm]{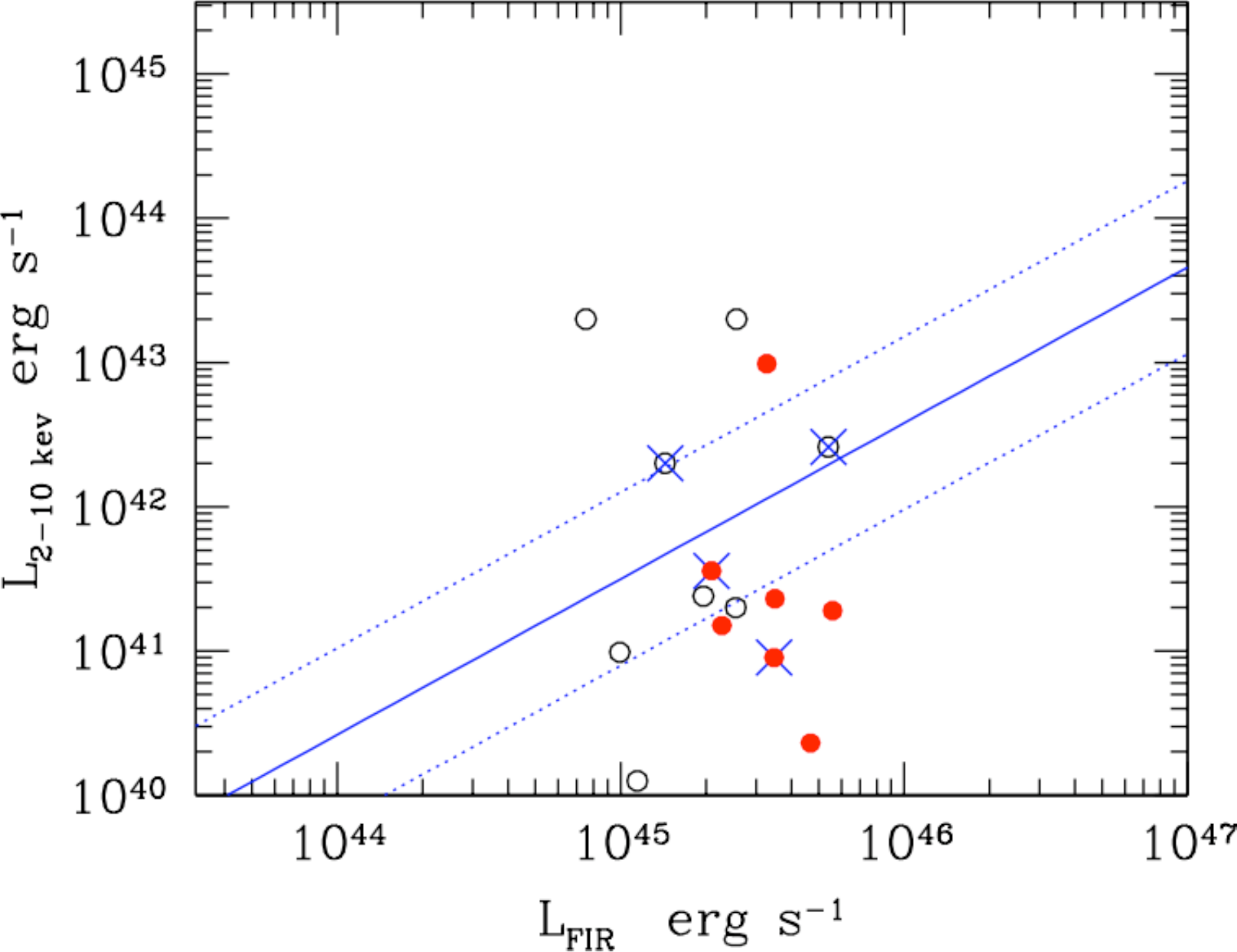} 
\caption{The X-ray (2-10 keV) vs. the FIR (4-100\,$\mu$m) relation for our
         sources. The solid and dotted lines denote the best-fit relation of
         \citet{Ranalli2003} - and its associated $2\sigma$ error, for normal
         galaxies in the local Universe. The red filled circles denote the
         sources which lie inside the box while open circles denote those outside 
          the box  in the diagnostic diagram of
         \citet{Papadopoulos2010}. Crosses denote the four Compton-thick
         sources.}
\label{ranalli}
\end{center}
\end{figure}

From Table 1 we see that a large number of sources (with the exception of
Mrk\,273, IRAS\,05189-2524 and the four Compton-thick AGN) are unobscured having
column densities $\rm N_H<10^{22}$\,\cunits. It is possible that in a number of
these sources the X-ray emission comes from star-forming processes rather than
the AGN. In order to explore this issue, we are plotting the observed
(uncorrected for obscuration) X-ray luminosity against the far-IR luminosity in
Figure\,\ref{ranalli}. \citet*{Ranalli2003} have found a strong correlation
between the X-ray and the far-IR luminosity for star-forming galaxies in the
local Universe. This implies that sources with an excess of X-ray emission
above this relation can be securely associated with AGN. The sources that fall
within the \citet{Ranalli2003} relation should be either `normal' galaxies
(i.e. without an AGN) or
highly obscured (Compton-thick) AGN. From Fig.\,\ref{ranalli} we see that three
sources (IRAS\,05189-2524, Mrk\,273 and NGC\,7469) present X-ray emission above that
of `normal' galaxies.

In order to further investigate the energy mechanism of the sources which lie
in the window proposed by P10, we use the mid-IR diagnostic diagram of
\citet{Laurent2000}. This traces the excess of the 5.5\,$\mu$m continuum
emission with respect to the 15\,$\mu$m continuum and 6.2\,$\mu$m PAH emission;
essentially it is a tracer of hot dust from the AGN which has a higher
5.5\,$\mu$m/15\,$\mu$m flux ratio than the cooler dust associated with
H{\scriptsize II} regions, and low 6.2\,$\mu$m PAH flux. The sources within 
the P10 box,  namely Mrk\,273,
IRAS\,09320+6134, IRAS\,10565+2448, Arp\,220, IRAS\,17208-0014, IRAS\,12112+0305,
IRAS\,23365+3604, show a fractional AGN contribution of 77\%, 80\%, 5\%, 60\%,
14\%, 21\% and 22\% respectively
\citep[][and V. Charmandaris private communication]{Armus2007}.
For comparison, we note that the two Compton-thick AGN
 outside the box  (Mrk\,231 and NGC\,6240) show AGN contributions of 95\% and 50\% 
 respectively \citep{Armus2007}.

\begin{table}
\centering
\caption{AGN diagnostics}
\label{agn_diagnostics}
\begin{tabular}{cccc}
\hline\hline
Name             & X-ray spectrum & X-ray - IR & mid-IR \\
 (1)             & (2)            & (3)        & (4)    \\
\hline
IRAS\,09320+6134 & \checkmark     &            & 80\%   \\
IRAS\,10565+2448 &                &            & 5\%    \\
IRAS\,12112+0305 &                &            & 21\%   \\
Mrk\,273         & \checkmark     & \checkmark & 77\%   \\
Arp\,220         & \checkmark     &            & 60\%   \\
IRAS\,17208-0014 &                &            & 14\%   \\
IRAS\,23365+3604 &                &            & 22\%   \\
 \hline\hline
\end{tabular}
\begin{list}{}{}
\item The columns are:
      (1) Name
      (2) AGN based on a flat or absorbed spectrum
      (3) AGN based on deviation from the \citet{Ranalli2003} relation
      (4) Contribution of an AGN to the MIR according to the \citet{Laurent2000}
          diagnostic
\end{list}
\end{table}

In Table\,\ref{agn_diagnostics} we summarize the results of the AGN diagnostics
used to examine the nature of the sources in the P10 box. There is an agreement
between the mid-IR and the X-ray diagnostics which find that 3/7 sources host
an AGN, if we use a 50\% as a threshold for the \citet{Laurent2000} diagnostic. 
We note that one source (IRAS\,12112+0305)
has been observed with the VLA \citep{Nagar2003}. This source, which is
optically classified as a LINER, appears to show a radio core at 15\,GHz with
a radio power of $\rm \log P\approx22.6\,W\,Hz^{-1}$. 
 Therefore, the possibility that this source hosts a Compton-thick AGN
  cannot be securely excluded.

\begin{figure}
\begin{center}
\includegraphics[width=9.cm]{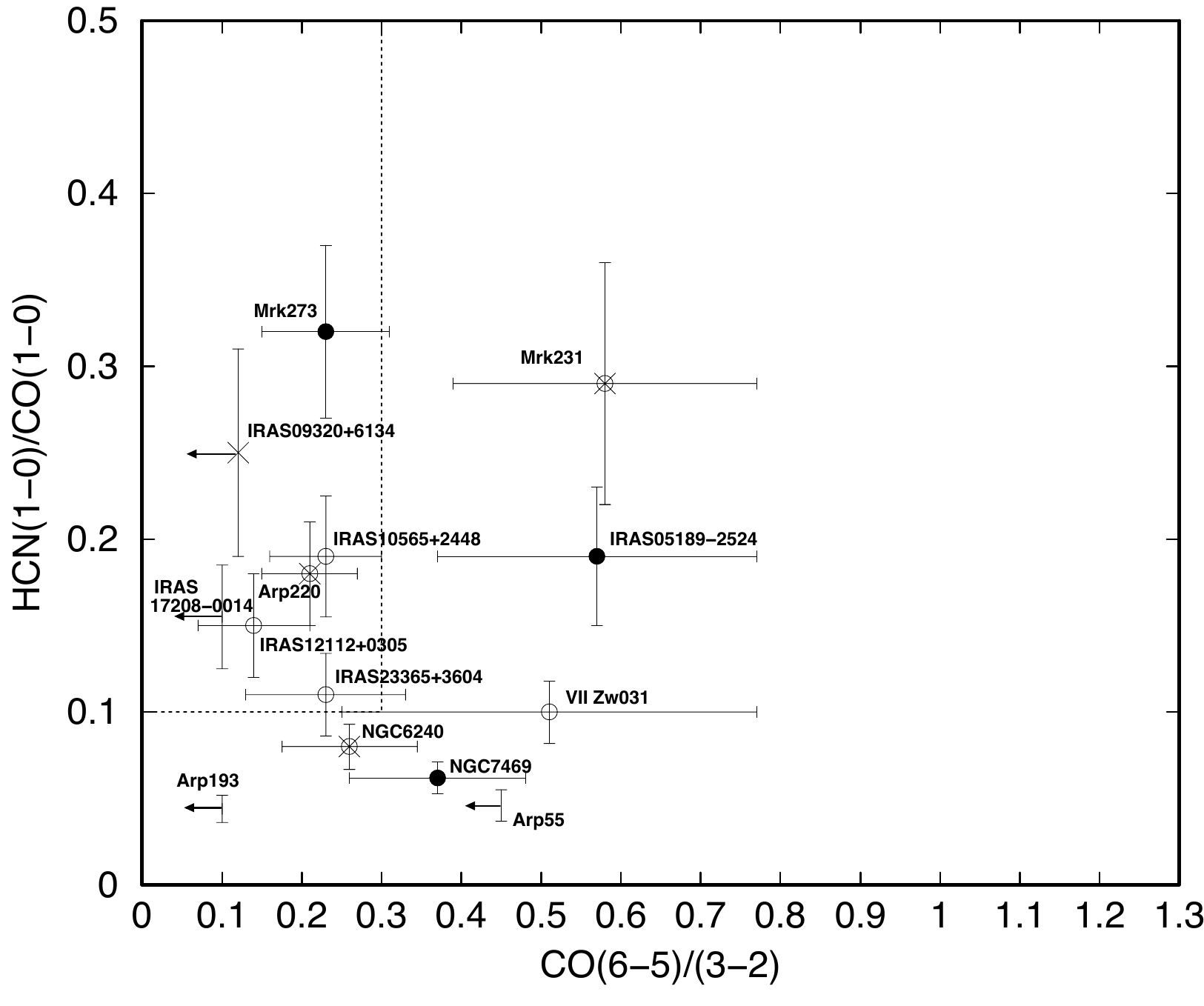} 
\caption{The diagnostic diagram of \citet{Papadopoulos2010}.
        High HCN(1-0)/CO(1-0) together with low CO(6-5)/(3-2) ratios
       (region inside the box) mark the region where large average
        dust absorption likely surpresses the high-J (in this case
        J=6-5) CO emission that is emitted by a heavily dust-enshrouded
       starburst with large supplies of dense gas (and hence the large
       HCN/CO J=1-0 ratio).
       Filled circles denote the excess X-ray emission sources (i.e.
      likely AGN)  according Figure\,\ref{ranalli}. Open circles denote
       sources falling within the \citet{Ranalli2003} relation for
       normal galaxies (all four CO(6-5)/(3-2) upper limits fall again
        within the same relation).  Crosses denote the four
       Compton-thick sources according to X-ray spectroscopy.}
\label{box}
\end{center}
\end{figure}

\section{Discussion \& Conclusions} 

In Figure\,\ref{box} we  present the diagnostic diagram of  P10.  We see
that out of the four  Compton-thick sources (marked with crosses), two
(IRAS\,09320+6134 and  Arp\,220) fall  within the `high  dust obscuration'
window. The  other one (NGC\,6240) lies  very close to the  box having a
marginally lower  HCN(1-0)/CO(1-0) ratio.  The  fourth source (Mrk\,231)
lies away  from the  box showing  highly excited CO  lines and  a high
HCN/CO J=1--0 line ratio. This presents no fundamental problem in
  the Compton-thick selection using molecular lines, as Compton-thick AGN
  are expected also to lie outside the  `window' of the diagnostic diagram of
  P10.  Moreover,  this source displays peculiar characteristics
relative to the  other three Compton-thick sources:  although  it is a
bona-fide Compton-thick  source, it is  a broad absorption  line (BAL)
QSO.  BAL is  a special  class of  QSOs that  show highly  ionized gas
flowing  away  from the  central  source  with  speeds in  some  cases
exceeding 30000\,$\rm km\,s^{-1}$. Most BAL QSOs present high obscuring
columns  in  X-rays  but  still their  broad-line-regions  are  viewed
relatively  unobscured  by dust,  probably  meaning  that their  dense
obscuring  medium is  dust-free \citep[e.g.][]{VeronCetty2000}.  It is
then reasonable  that in the  case of Mrk\,231 the  high CO(6-5)/CO(3-2)
ratio implies that the CO  emission is not suppressed by dust, despite
the presence of a large X-ray obscuring column density.
 
In addition,  the question that needs  to be addressed is  what is the
nature of the remaining sources within the P10 `high dust obscuration'
box.  Apart  from the  two Compton-thick AGN  there are  five more
sources presenting characteristically low CO(6-5)/CO(3-2) ratios in
combination  with   high  HCN/CO  J=1-0  ones.    These  are  Mrk\,273,
IRAS\,17208-0014,  IRAS\,10565+2448,  IRAS\,23365+3604,  and  IRAS\,12112+0305.
 Among  them only  Mrk\,273  has  a  known X-ray  luminous,  highly
  obscured   ($\rm  N_H\sim   4\times  10^{23}$\,\cunits),   but  not
Compton-thick AGN.  
Thus total mean column density of a gaseous disk around the AGN
that would be responsible for its X-ray obscuration would be
$\rm <N_H(disk)> \sim  2 xN_H = 8 \times 10^{23} cm^{-2}$. On the other hand mean
Êdisk column densities as low as $\rm <N_H(disk)>=5.2 \times 10^{24} cm^{-2}$ are
compatible with the observed suppression of the CO 6-5/3-2 ratio in
this object, which are
6.5 times higher than the one deduced by the X-ray data. This discrepancy,
Êwhile not very large given the totally independent means used to obtain
$\rm <N_H(disk)>$ (molecular lines vs X-rays), remains considerable.  
  However,  we note that  the large difficulty of  measuring CO
  J=6-5 fluxes from  the Earth, with a large  telescope like the JCMT,
  will more  often yield underestimates rather  than overestimates (as
  pointing  errors become  substantial  with respect  to narrow  beams
  used).  This  could tend  to make  molecular  line-deduced and
  X-ray-deduced hydrogen  column densities more  compatible in cases
  like Mrk\,273.

The  X-ray  emission of  the  remaining four  sources  in  the box  is
probably dominated by star-forming  processes according to their X-ray
spectra, the $\rm L_X-L_{FIR}$ relation and the MIR diagnostics of
\citet{Laurent2000}. This could explain  why the sources do  not appear as
Compton-thick in X-ray wavelengths, as the origin of the X-rays is
  not a potentially deeply buried  AGN but stellar.  Thus such sources
  are not  useful in deducing  mean column densities of  gaseous disks
  around  AGN  (and  could   be  indeed  found  inside  the  selection
  ``window'' of P10 without invalidating its utility).

 As a  consistency  check it  is  interesting to  discuss the  Si
  $\rm 9.7\,\mu m$  properties of the  objects within the  P10 box, as  it is
  clearly impossible that these  have  substantial dust optical depths in short
  sub-mm wavelengths as  suggested by P10,  while at the same time 
   they exhibit modest or low
  Si (9.7$\mu m$) optical depths \citep[e.g.][]{Shi2006}.  Therefore
the  highly obscured  objects according  to P10  should present
large  Si   optical  depths.   There   are  {\it Spitzer}-IRS  spectroscopic
observations for four sources  in the P10 box: Mrk\,273, IRAS\,09320+6134,
Arp\,220 and IRAS\,12112+0305 \citep{Armus2007}.  Their Si ($\rm 6.7\,\mu m$)
optical  depths are  high: 1.8$\pm$0.4,  1.6$\pm$0.3,  3.3$\pm$0.2 and
1.3$\pm$0.3 respectively. This clearly  suggests an association of all
the sources in the P10 box with optically-thick systems at $\rm 9.7\,\mu m$.

In conclusion, our results show that the  diagnostic proposed by
  P10 offers  a potentially  useful tool for  the detection  of highly
  obscured X-ray  luminous AGN . This bears important implications 
   for future facilities such as ALMA. However, our work identified 
    two caveats. First, the AGN inside the P10 box 
     are heavily obscured but not necessarily Compton-thick,
      as the case of Mrk273 implies. Second, there can be large 
       contaminations by powerful star-forming systems.  
  Additional X-ray
  spectral observations  might be  necessary in  order to  sift those  out. 
    More molecular line  data, ranging
  from  low to high-J  CO lines,  as well  as low-frequency  (and thus
  unaffected   by   dust  exinction)   density   indicators,  such   as
  low-to-modest-J  HCN and  CS  lines of  nearby  (U)LIRGs are  needed
  (along with  X-ray data)  to verify (or  not) this picture  in local
  (U)LIRGs  that  are known  to  have  X-ray  luminous AGN.   The  now
  superbly  operating  {\it Herschel  Space  Observatory}  that  can  detect
  numerous  high-J CO  lines in  local  (U)LIRGs \citep[e.g.][]{vanderWerf2010}
{\it in  combination} with ground based low-J  CO and HCN, CS
  observations presents  an excellent opportunity to do  so. The large
  (U)LIRG  sample now  being  observed  as part  of  the {\it Herschel}  Key
  Project  HerCULES,  in combination  with the
  extensive  ground-based  low  frequency  CO  and  HCN  survey  will be ideal for such studies.

\begin{acknowledgements} 
We would like to thank the anonymous referee for many useful comments and suggestions which helped to improve 
 substantially this paper.
IG acknowledges the receipt of a Marie Curie fellowship FP7-PEOPLE-IEF-2008 Prop. 235285. We
thank P. P. Papadopoulos for reading the manuscript and providing useful
comments and suggestions. The \chandra data were taken from the \chandra Data
Archive at the \chandra X-ray Center. 
\end{acknowledgements}

\end{document}